\documentclass[]{aa}
\usepackage[latin1]{inputenc}
\usepackage[english]{babel}
\usepackage{amsmath}
\usepackage{amsfonts}
\usepackage[]{bigints}
\usepackage{amssymb}
\usepackage[]{graphicx}
\usepackage{multirow}
\usepackage[]{epstopdf}
\usepackage{natbib,twoopt}
\usepackage[breaklinks=true]{hyperref} 
\bibpunct{(}{)}{;}{a}{}{,} 
\makeatletter

\title{High-contrast imaging in polychromatic light with the self-coherent camera} 

\titlerunning{High contrast imaging in polychromatic light with the SCC}
\authorrunning{J. Mazoyer et al.}

\author{J.~Mazoyer\inst{\ref{inst1}}\and P.~Baudoz\inst{\ref{inst1}}\and
R.~Galicher\inst{\ref{inst1}}\and G.~Rousset\inst{\ref{inst1}}}

\institute{LESIA, Observatoire de Paris, CNRS, UPMC Paris 6 and Denis Diderot Paris 7, Meudon, France. \email{johan.mazoyer@obspm.fr}\label{inst1} 
}


\abstract 
{In the context of direct imaging of exoplanets, coronagraphs are commonly proposed to reach the required very high contrast levels. However, wavefront aberrations induce speckles in their focal plane and limit their performance.}
{An active correction of these wavefront aberrations using a deformable mirror upstream of the coronagraph is mandatory. These aberrations need to be calibrated and focal-plane wavefront-sensing techniques in the science channel are being developed. One of these, the self-coherent camera, of which we present the latest laboratory results.}
{We present here an enhancement of the method: we directly minimized the complex amplitude of the speckle field in the focal plane. Laboratory tests using a four-quadrant phase-mask coronagraph and a 32x32 actuator deformable mirror were conducted in monochromatic light and in polychromatic light for different bandwidths.}
{We obtain contrast levels in the focal plane in monochromatic light better than $3.10^{-8}$ (RMS) in the 5 -- 12 $\lambda/D$ region for a correction of both phase and amplitude aberrations. In narrow bands (10 nm) the contrast level is $4.10^{-8}$ (RMS) in the same region.}
{The contrast level is currently limited by the amplitude aberrations on the bench. We identified several improvements that can be implemented to enhance the performance of our optical bench in monochromatic as well as in polychromatic light.}

\begin{document}
\maketitle

\section{Introduction}
\label{sec:intro}

Planets detected with indirect methods often orbit their star at a small semi-major axis, while direct imaging targets objects located at larger distances from their star. Therefore, the development of this technique might lead to the discovery of entirely new objects, following the detection of $\beta$-pic b \citep{Lagrange_Bpic09} and of the HR~8799 system \citep{Marois08,Marois10}. Direct imaging performed on broad bandwidths will also allow surface and atmosphere spectroscopic analysis. However, the main challenge of direct imaging is the high contrast level required to separate the planet signal from the stellar light.

Most direct-imaging instruments use coronagraphs to reject the stellar light. But their performance is strongly limited by phase and amplitude aberrations in the incident wavefront, which produce stellar speckles in the image plane. To compensate for all these aberrations and reduce the speckle level, an active correction including a focal plane wavefront sensor paired with a deformable mirror (DM) is mandatory. Several techniques have been developed to retrieve phase and amplitude aberrations directly in the focal plane of a coronagraph, using either the application of known phases on the DM \citep{BordeTraub06} or a specific instrumental design. Among these, the self-coherent camera \citep[SCC, ][]{Baudoz06} uses the coherence between the focal plane speckles and the stellar light rejected by the coronagraph. This technique has already reached high contrast levels in simulations \citep{Galicher10} and experimentally \citep{Mazoyer13} in monochromatic light. 

This paper presents the latest laboratory performance obtained with the SCC. An enhancement of the method is described in Section~\ref{sec:theo_part}, and we present the current bench configuration in Section~\ref{sec:setup}. Results are presented in Section~\ref{sec:monochromatic} for monochromatic light and in Section~\ref{sec:polyc} for polychromatic light.

\section{Speckle minimization}
\label{sec:theo_part}

\begin{figure}[]
 \begin{center}
  \includegraphics[height=2.4cm]{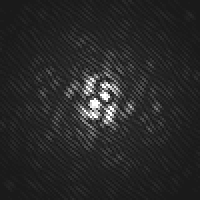}
  \includegraphics[height=2.4cm]{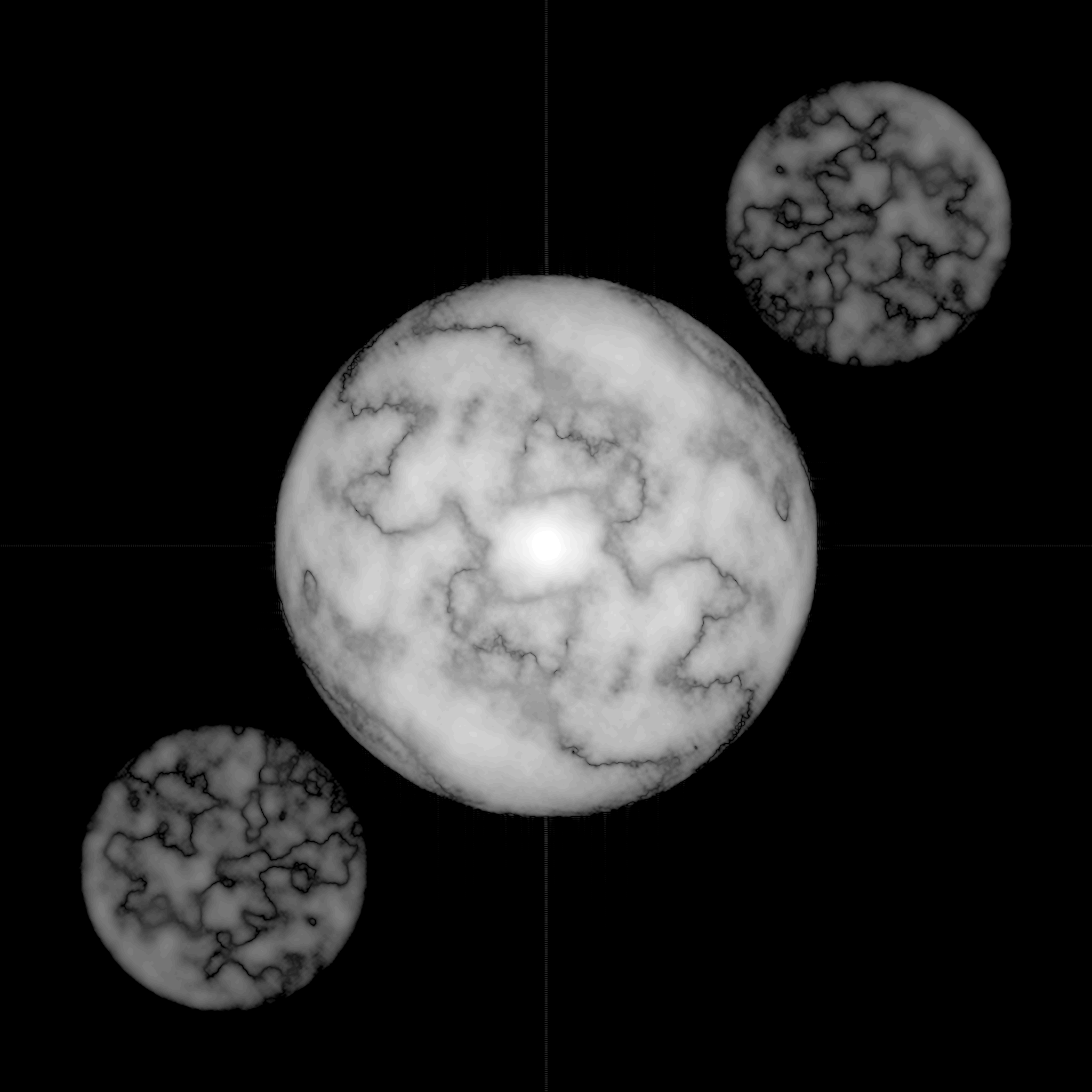}
\end{center}
 \caption[SCC_proc] 
{ \label{fig:SCC_proc} Intensity of the speckle field (simulation) in the focal plane of an SCC (left) and its Fourier transform (right) for monochromatic light.}
\end{figure}

In this section, we briefly recall the formalism used in \cite{Mazoyer13} and introduce the improvement added to this technique. In the SCC, we modify the Lyot stop of a coronagraph by adding an off-axis small circular reference hole. This hole selects part of the coherent light rejected by the coronagraphic mask and creates interference fringes over the speckle field in the focal plane (Figure \ref{fig:SCC_proc}, left). We call $\vec{\xi_0}$ the 2D-position of the reference hole relative to the center of the pupil. $A_S$ and $A_R$ are the complex amplitudes in the focal plane of the speckles and of the reference wave. The intensity $I$ in the SCC image at a given wavelength $\lambda$ is
 \begin{equation}
\label{eq:I_monoc_1ref}	
\begin{array}{c}
\ I(\vec{x}) = |A_{S}(\vec{x})|^2 + |A_{R}(\vec{x})|^2 + \\
\ A_{S}(\vec{x})A_{R}^{*}(\vec{x})\exp\left(\frac{2i\pi\vec{x}.\vec{\xi_0}}{\lambda} \right) +\\
\ A_{S}^{*}(\vec{x})A_{R}(\vec{x}) \exp\left(-\frac{2i\pi\vec{x}.\vec{\xi_0}}{\lambda} \right),
 \end{array}	
	\end{equation}
where $\vec{x}$ is the coordinate in the focal plane and $A^{*}$ the conjugate of $A$. The last two terms in Equation~\ref{eq:I_monoc_1ref} express the spatial modulation with a fringe period of $\lambda/\Vert\vec{\xi_0}\Vert$. We call $D$ and $D_R$ the diameters of the Lyot and reference holes. If $\Vert\vec{\xi_0}\Vert > 1.5 D$, the Fourier transform ($\mathcal{F}$) of I (represented in Figure \ref{fig:SCC_proc}, right) is composed of a central lobe (Fourier transform of the first line in Equation \ref{eq:I_monoc_1ref}) and two lateral lobes, $\mathcal{F}[I_{+}]$ and $\mathcal{F}[I_{-}]$ (two other lines). We select $\mathcal{F}[I_{-}]$, center it, and apply an inverse Fourier transform to obtain $I_-$, which we use to minimize the speckle field:
\begin{equation}
	\label{eq:estimator}
I_{-} = A_{S}A_{R}^{*}.
\end{equation}

The largest correction zone in the focal plane achievable by an NxN actuator DM is the square $[-N,N]^2$ in $\lambda/D$. We refer to this zone as the dark hole (DH). If the reference hole is small enough \citep[$D_R < \frac{1.22\sqrt{2}}{N}D$, ][]{Mazoyer13}, then $A_{R}^{*}$ is nonzero over the DH and all speckles are fringed in the DH. Under this assumption, the minimization of $I_-$ corresponds to the minimization of the fringed speckles in the full DH.

In previous papers \citep{Galicher10,Mazoyer13}, $I_-$ was used to retrieve the phase and amplitude aberrations in the pupil plane upstream of the coronagraph and minimize them. This method was dependent on a model of the coronagraph. In this letter, we directly minimize $I_-$ because our goal is to minimize the complex amplitude of the fringed speckles in the focal plane, not the aberrations in the pupil plane. For that purpose, we used a linearized approach technique as in classical adaptive-optic systems \citep{Boyer90}. We used a modal actuator basis composed of sinus and cosinus functions for all the spatial frequencies achievable by the DM \citep{Poyneer05}. This basis is interesting because it creates localized speckles in the focal plane \citep{MazoyerSPIE13}. For each mode applied to the DM, we recorded the corresponding complex quantity $I_-$. We built the interaction matrix that links the actuator voltages to their signature in $I_-$. The generalized inverse of this matrix by the singular-value decomposition (SVD) method allowed us to compute the control matrix, which gives the DM surface $\phi_{DM}$ that minimizes any measured $I_-$ in the least-squares meaning. This matrix can be applied in a closed loop between the focal plane and the DM. We recalibrated this matrix after achieving a first deep correction to improve the contrast level.

\section{Experimental setup}
\label{sec:setup}

Our high-contrast bench is described in \cite{MazoyerSPIE13}. In this section, we briefly present the main parameters used in the current experiments. The diameters of the pupils are 8.1 mm for the unobstructed entrance pupil, 8 mm for the Lyot pupil, and $0.35$ mm for the reference hole. We set $\Vert\vec{\xi_0}\Vert = 1.8D$.

The focal plane coronagraph mask is a four-quadrant phase mask \citep[FQPM, ][]{Rouan00}. This coronagraph mask induces a $\pi$ phase shift in two quadrants in diagonal with respect to the two other quadrants. In our experiment, the $\pi$ phase shift is induced by a step in the material and is therefore optimized for a single wavelength of $637$ nm ($\pm 5$ nm). 

We used an optical fiber source fed either by a monochromatic laser diode or by a filtered supercontinuum laser source. Table~\ref{tab:filtres} presents the central wavelengths ($\lambda$) and spectral bandwidths and resolutions ($\Delta\lambda$ and $R$) for all the spectral filters applied to the white source and for the monochromatic diode.
\begin{table}[ht]
\center \begin{tabular}[c]{|c|c|c|c|}
\hline 
Source & $\lambda$ (nm) & $\Delta\lambda$ (nm) & $R$ \\ 
\hline 
Diode & 636.9 & $<$ 1 & $>600$ \\ 
\hline 
 & 633.2 & 7.5 & 84.4 \\ 
\cline{2-4} 
White source &637.4 & 9.9 &  64.4  \\ 
\cline{2-4} 
&643.7 & 9.4 & 68.5 \\ 
\cline{2-4} 
&652.2 & 35.0 & 18.6 \\ 
\cline{2-4} 
&657.1 & 8.6 & 76.4 \\ 
\hline 
\end{tabular} 
\caption{Central wavelengths and bandwidths of the filters.}
{ \label{tab:filtres}}
 \end{table}

Out of the 32x32 actuators of the DM, we used 27 actuators across the pupil ($N = 27$). As explained in \cite{BordeTraub06}, phase and amplitude aberrations correction is possible with one DM on a half-DH. We also shrank the correction zone of 1 $\lambda/D$ to enhance performance. Finally, the actual correction zone was $[0,13]\times[-13,13]$ (in $\lambda/D$ at 637 nm). The contrast levels were retrieved by normalizing by the highest value of noncoronagraphic image for the same source intensity. In the next sections, we successively show results obtained on the optical bench in monochromatic and polychromatic light.

\section{Monochromatic light results}
\label{sec:monochromatic}

\begin{figure*}
\begin{center}
\begin{tabular}{c}
\includegraphics[height=6.4cm]{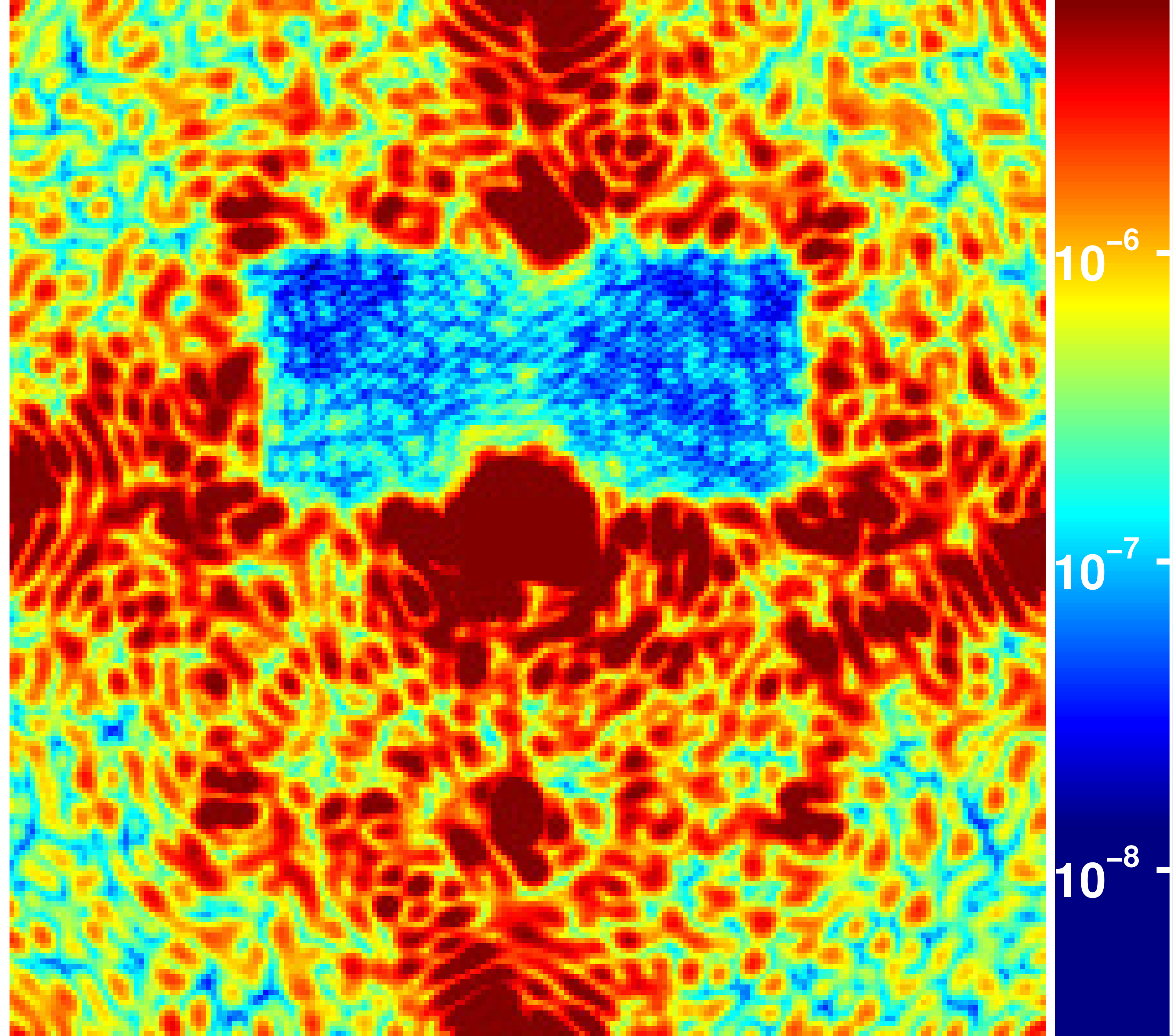}
\includegraphics[trim=1.3cm 0.6cm 1.1cm 1.2cm, clip=true, height=6.4cm]{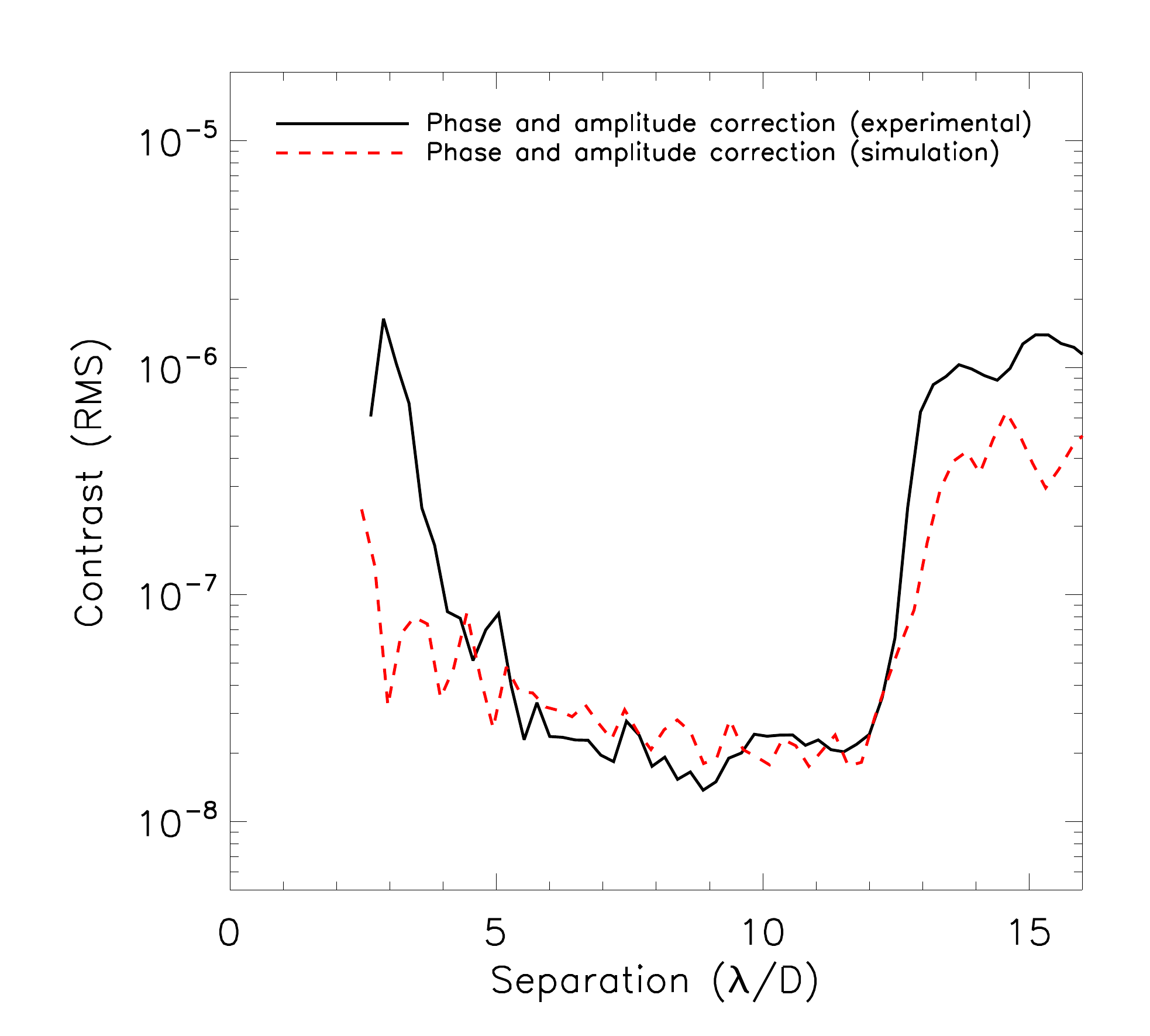}
\end{tabular}
\end{center}
\caption[res_mono]
{\label{fig:res_mono} Left: experimental half DH obtained in monochromatic light. Right: radial profiles obtained for phase and amplitude correction on the experimental bench (black solid line) and in simulation with realistic amplitude aberration and no phase errors (red dashed line).}
\end{figure*}

The DH obtained in the focal plane after correction in monochromatic light is presented in Figure~\ref{fig:res_mono} (left). High-contrast imaging was performed, and we obtained a half-DH. In this DH, correction levels can reach levels better than $10^{-8}$. However, we can see that speckles near the FQPM transitions (vertical and horizontal axis) are brighter than those in other parts of the DH. The phase of the electrical field shift quickly of $\pi$ when a transition is crossed and a misregistration due to movement of the camera has more impact on the correction efficiency. However, the image of a putative planet located on a transition would be distorted and strongly attenuated, and the contrast level in these regions is not fully relevant. Therefore, we excluded 1.8 $\lambda/D$ around the vertical transition and 1.8 $\lambda/D$ above the horizontal transition in measuring the radial profile of the azimuthal standard deviation (in RMS) of the contrast levels. We plot this profile as a function of the distance to the star (in $\lambda/D$) as a black solid line in Figure~\ref{fig:res_mono} (right). The contrast levels obtained are better than $3.10^{-8}$ in the 5 -- 12 $\lambda/D$ zone.

These results are better by a factor 13 than the results presented in \cite{Mazoyer13} on the same bench for two reasons. First, this improvement is mainly due to the fact that no assumption on the coronagraphic mask or on the reference hole is made to retrieve the complex field associated to the speckles, whereas the estimation in \cite{Mazoyer13} was model-dependent. The use of a new detector with a better dynamic range also limits the saturation in the image (saturation creates oscillations in the Fourier plane), which allows a more stable and deeper correction.

At the end of the correction, speckles are still fringed in the DH (Figure~\ref{fig:res_mono}, left), which means that some speckles are currently encoded by the SCC but beyond correction capability. Most of the amplitude aberrations (which are about 10\% RMS in intensity) are induced by a vignetting effect linked to high-frequency structures in the DM surface \citep[see ][]{Mazoyer13}. We simulated an SCC correction introducing amplitude-only aberrations measured on the bench (Figure 11 \textit{ibid.}). The radial profile of the result image of the simulation is plotted in Figure~\ref{fig:res_mono} (right) as a red dashed line. In the 3 -- 13 $\lambda/D$ region, the simulation (without phase error or noise) and experimental results match well, which shows that the residuals after correction are set by the high-amplitude errors, not by the SCC performance. Indeed, even if we correct on a half-DH, the speckles in the uncorrected half-DH are leaking into the corrected area \citep{Giveon06}. This limit may be overcome by introducing a second DM on the optical bench \citep{Pueyo10}. 

Simulation and experimental results differ below 3 $\lambda/D$ and outside the DH ($>$ 13 $\lambda/D$). The differences below 3 $\lambda/D$ can be explained by a possible poor estimation and correction of the low-order amplitude aberrations due to the saturation of the detector in the center. Outside the DH, phase aberrations are not corrected, thus simulation (without phase error) and experimental correction (with phase errors) are not expected to match. 

\section{Polychromatic light results}
\label{sec:polyc}

The FQPM used on the bench limits the bench performance in polychromatic light. Indeed, the phase shift induced by a monochromatic FQPM is slightly different for each wavelength \citep{Riaud03}. Therefore, the on-axis light at a wavelength different from the optimal one is not completely rejected outside the Lyot pupil. This light creates an Airy pattern \citep{Galicher11} in the image plane after the Lyot, which, because it is not created by wavefront errors, is scarcely corrected by the DM.

\begin{figure*}
\begin{center}
\begin{tabular}{c}
\includegraphics[trim=1.3cm 0.6cm 1.1cm 1.2cm, clip=true, height=6.4cm]{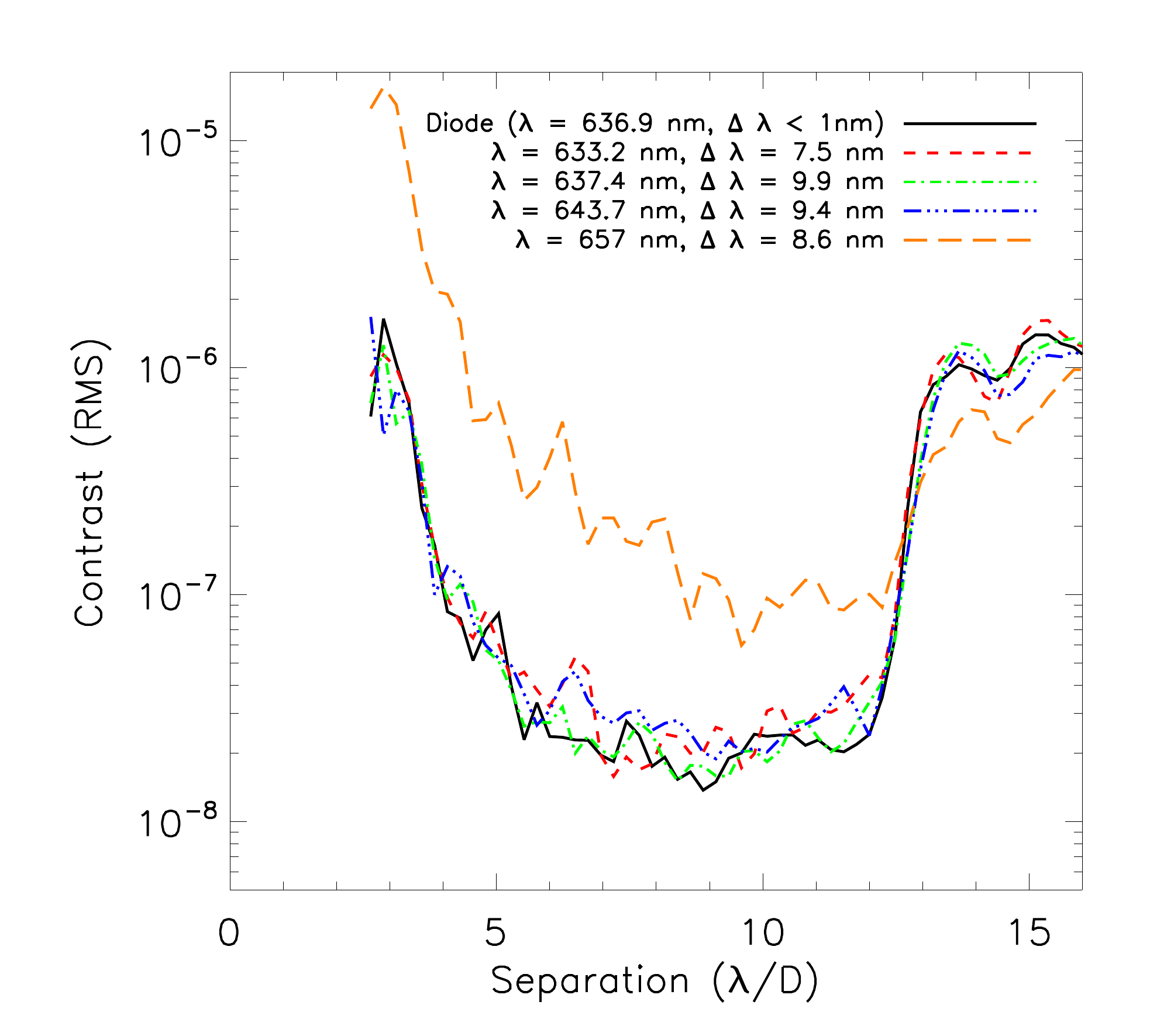}
\includegraphics[height=6.4cm]{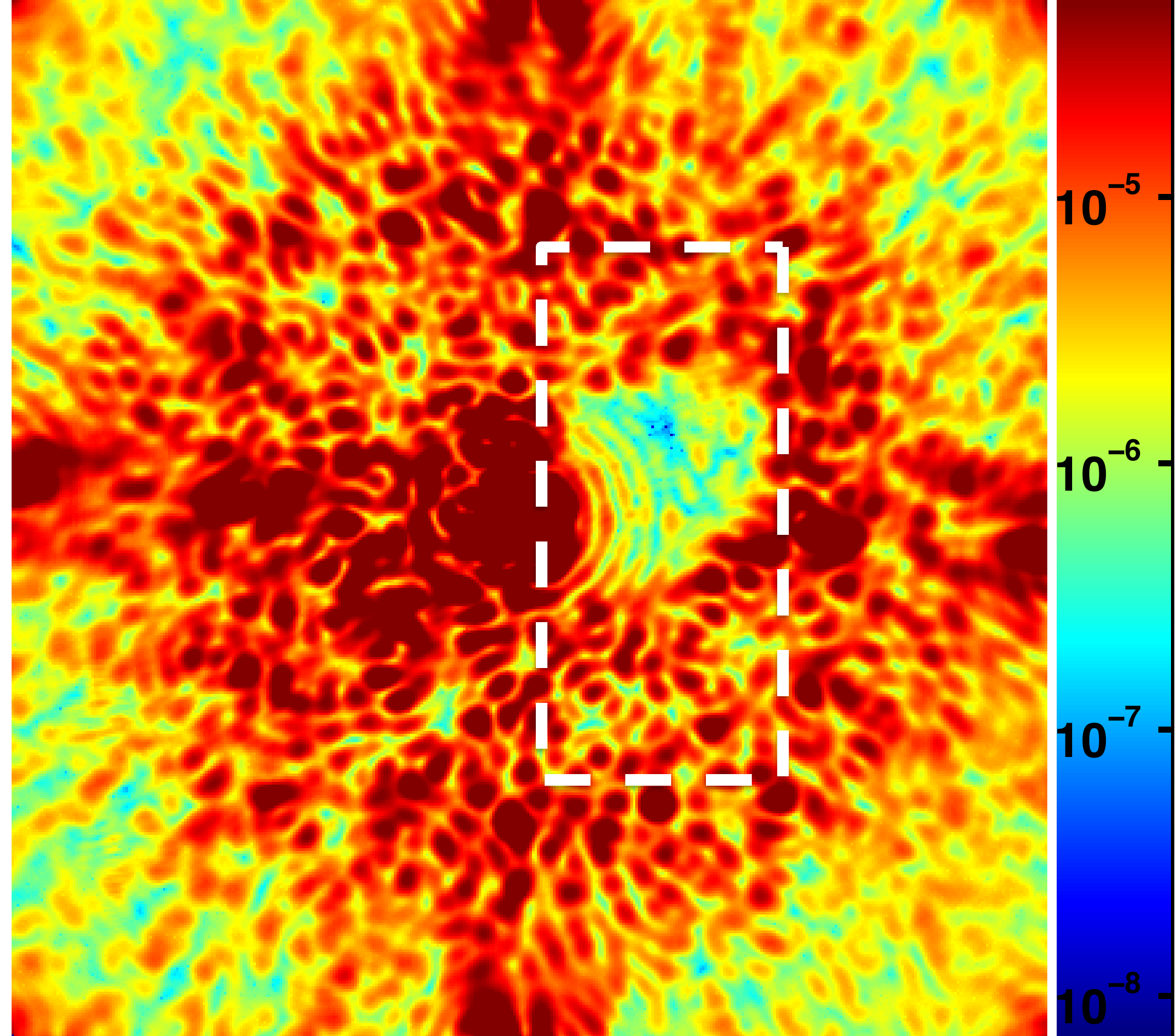}
\end{tabular}
\end{center}
\caption[res_poly]
{\label{fig:res_poly} Left: radial profiles of the azimuthal standard deviation (RMS) of the contrast levels obtained experimentally for all filters for phase and amplitude correction. Right: half-DH for $\lambda = 652.2$ nm and $\Delta\lambda = 35.0$ nm.}
\end{figure*}

\cite{Galicher10} also pointed out specific problems in the SCC estimation at broad bandwidths. In addition to the smearing of the speckles, the interference fringes have wavelength-dependent periods ($\lambda/\Vert\vec{\xi_0}\Vert$). This results in a blur of the fringes far away from the center of the image. Therefore, the fringes are contrasted only inside a stripe and the correction is hence limited to this zone. Assuming a flat spectrum, the width of this stripe is $R\lambda/(2\Vert\vec{\xi_0}\Vert)$. The chromaticity can therefore limit the correction, according to whether this width is broader than the DH size or not. The stripe covers the whole DH if, and only if,
\begin{equation}
\label{eq:min_res}
R \gtrsim N\sqrt{2}\frac{\Vert\vec{\xi_0}\Vert}{D},
\end{equation}
assuming that the fringe of null optical path difference is centered on the image. For example, to cover at least a DH of $[-13,13]^2$ (in $\lambda/D$) with $\Vert\vec{\xi_0}\Vert = 1.8D$, Equation~\ref{eq:min_res} leads to $R \gtrsim 66$. 
We present here the correction for speckles in polychromatic light, first in narrow bandwidths (when Equation~\ref{eq:min_res} is satisfied), then for $R = 18.6$, where the correction is expected to be limited by the chromaticity of the SCC.

To obtain corrections in polychromatic light, we created an interaction matrix for each filter using the method presented in Section~\ref{sec:theo_part}. These matrices were then used to perform active correction on half-DH at each filter. The radial profiles of these results as a function of the distance to the star are plotted in Figure~\ref{fig:res_poly} (left). The abscissa is in $\lambda/D$  at 637 nm.

When using the filters centered on $\lambda = 633.2$ nm, $\lambda = 637.4$ nm and $\lambda = 643.7$ nm, the results do not differ strongly from the monochromatic case. For these wavelengths, which are close to the optimal wavelength of the coronagraphic mask, contrasts are better than $3.8 .10^{-8}$ in the 5 -- 12 $\lambda/D$ zone. The performance with the $\lambda = 657.1$ nm filter is more limited than the monochromatic case: a central Airy pattern appears in the center of the focal plane, because the light is only poorly rejected by the FQPM. Because they is not created by aberrations, these chromatic coronagraphic leaks can scarcely be corrected by the DM.

We also performed a correction in a half-DH with the $R = 18.6$ filter. In this case, Equation~\ref{eq:min_res} is not satisfied and the correction is only achieved on a stripe. In Figure~\ref{fig:res_poly} (right), we represent the correction on a quarter-DH, superposition of a half-DH (shown as the white dashed rectangle) and the stripe. In this DH, the results can reach $10^{-7}$. However, we can see the wings of the Airy pattern cause by FQPM chromatic leaks, which severely limit the performance in contrast.

Several solutions have been investigated to enhance the correction zone \citep{Galicher10}. The association of an SCC with a Wynne compensator would prevent the blurring of the fringes. We might also use an integral field spectrometer (IFS), which is commonly used in high-contrast instruments for the detection and analysis of exoplanets. In each channel, if the bandwidth meets the requirements of Equation~\ref{eq:min_res}, fringes would be contrasted in the full DH and an estimation of the speckle complex field would be possible. Finally, a slight modification of the SCC design might enhance the correction zone despite the blurring: the addition of a second reference hole in the Lyot plane produces another fringe pattern. If these two fringe patterns are oriented in a different direction, any speckle would be fringed by at least one of the reference wave, which would allow a correction on a larger zone.

\section{Conclusion}
\label{sec:ccl}

We presented the latest improvements made to the self-coherent camera (SCC). We tested this technique on a laboratory bench and reached contrast levels (RMS) better than $3.10^{-8}$ in the 5 -- 12 $\lambda/D$ zone. This means that we improved the contrast level by a factor of 13 with respect to the previous version of the SCC~\citep{Mazoyer13}. Numerical simulations showed that the current limitation is imposed by uncontrolled amplitude defects. For wavelengths close to the optimal wavelength of the coronagraphic mask and for narrow bands, contrast levels measured in the laboratory are better than $4.10^{-8}$ in the 5 -- 12 $\lambda/D$ zone. The current limitation is the chromaticity of our coronagraph. As expected, in broader bandwidths, a dark hole can be obtained in a smaller correction zone with degraded performance.

To enhanced the size and contrast level of this zone, we will study the multireference SCC, including experimental results, in a forthcoming paper. We also plan to use of the SCC with more achromatic coronagraphs, such as the multi-FQPM \citep{Galicher11} or the dual-zone phase mask \citep{Ndiaye12}. This is expected to improve the chromatic behavior and show that the SCC is compatible with several coronagraphic masks.

\bibliographystyle{aa} 
\bibliography{bib_Maz2013}   

\begin{acknowledgements}
J. Mazoyer is grateful to the CNES and Astrium (Toulouse, France) for supporting his PhD fellowship.
 \end{acknowledgements} 
 
\end{document}